\documentclass[reprint,prl,twocolumn,superscriptaddress,nofootinbib,floatfix]{revtex4-1}
\usepackage{graphicx,bm,color,amsmath,amssymb}

\newcommand{\ba}{\begin{eqnarray}}
\newcommand{\ea}{\end{eqnarray}}
\newcommand{\be}{\begin{equation}}
\newcommand{\ee}{\end{equation}}
\newcommand{\bi}{\begin{itemize}}
\newcommand{\ei}{\end{itemize}}
\newcommand{\non}{\nonumber\\}
\newcommand{\D}{\nabla}

\begin{document}

\title{Criteria for resolving the cosmological singularity in Infinite Derivative Gravity around expanding backgrounds}
\author{James Edholm}
\author{Aindri\'u Conroy}
\affiliation{Consortium for Fundamental Physics, Lancaster University, Lancaster, LA1 4YB, United Kingdom}

\begin{abstract}
We derive the conditions whereby null rays `defocus' within Infinite Derivative Gravity for perturbations
around an (A)dS background, and show that it is therefore possible to avoid singularities within this framework. This is in contrast to Einstein's theory of General Relativity. We further extend this to an (A)dS-Bianchi I background metric, and also give an example of a specific perturbation where defocusing is possible given certain conditions.
\end{abstract}
\maketitle
\section{Introduction}
Einstein's theory of General Relativity (GR) has had many successes in the infrared regime (IR)~\cite{Will}, 
but it breaks down in the ultraviolet (UV) regime due to its admittance of black hole and cosmological singularities~\cite{Hawking:1973uf,Raychaudhuri:1953yv}.

%String-inspired ghost-free 
 Initial attempts to resolve these problems by introducing higher derivative terms into a gravitational action, such as Stelle's 4th derivative theory~\cite{Stelle:1976gc}, 
faltered due to the Ostrogradsky instability~\cite{Barnaby:2007ve}. 
When interactions take place in such a system, the vacuum decays into both positive and negative states and is therefore unstable~\cite{VanNieuwenhuizen:1973fi,Woodard:2015zca,Ostrogradsky:1850fid}. 
However this instability is avoided for infinite derivative theories of gravity (IDG). The infinite number of derivatives allows for the propagator to be modified in such a way
 that neither unwanted additional poles are introduced nor negative residues in the form of ghosts.
Infinite derivative actions, as seen in string theory~\cite{Tseytlin:1995uq}, were first applied to gravity~\cite{Biswas:2005qr} and investigated around a Minkowski background~\cite{Biswas:2011ar}
and later around backgrounds of constant curvature, such as (Anti) de Sitter space~\cite{Biswas:2016etb}. 

It is well known that GR contains a single (benign) pole in the graviton propagator. To avoid introducing any further poles  we can demand 
that any infinite derivative function appearing in the denominator of the propagator is the \textit{exponential of an entire function} of the d'Alembertian operator
$\Box=g^{\mu\nu} \nabla_\mu \nabla_\nu$, which by definition contains no 
zeroes~\cite{Tomboulis,Siegel:2003vt,Biswas:2005qr,Biswas:2011ar,Biswas:2013kla,Biswas:2016etb,Buoninfante:2016iuf}. IDG is modulated by a mass scale $M$, 
which determines the length scales below which the infinite derivative terms start to have an effect.
 
It has been shown that the propagator for IDG is renormalisable for higher orders~\cite{Talaganis:2014ida,Modesto:2011kw,Modesto:2012ys,Modesto:2014lga,Modesto:2016max}.
A non-singular  solution was constructed at the linearised level for 
both a static background~\cite{Biswas:2005qr,Biswas:2011ar,Edholm:2016hbt,Conroy:2017nkc,Conroy:2014eja} and a time-dependent background~\cite{Frolov,,Frolov2015mfb,Frolov2015scs} within IDG. 
The full equations of motion for this theory were found in~\cite{Biswas:2013cha}. Bouncing cosmological solutions~\cite{Calcagni:2013vra}, 
making use of the Ansatz $\Box R=c_1 R + c_2$, were examined in~\cite{Biswas:2010zk,Biswas:2012bp,Koshelev:2012qn,Koshelev:2013lfm}.

Recent work has focused on finding the quadratic variation of the action~\cite{Biswas:2016etb,Biswas:2016egy};
the Wald entropy~\cite{Conroy2015wfa}; the boundary terms~\cite{Teimouri:2016ulk};  the Hamiltonian~\cite{Mazumdar:2017kxr}; a black hole solution~\cite{Modesto:2010uh};
 radiation emission~\cite{Frolov:2016xhq}; using IDG as an EFT for M-theory~\cite{Calcagni:2014vxa}; the bending of light near the Sun~\cite{Feng:2017vqd};  the diffusion equation~\cite{Calcagni:2010ab};
black hole event horizons~\cite{Koshelev:2017bxd} and the stability of the Schwarzschild solution~\cite{Calcagni:2017sov}; IDG's effects on inflation~\cite{Biswas:2012bp,ArkaniHamed:2002fu} and in particular perturbations in the early universe using Cosmic Microwave Background data~\cite{Edholm:2016seu,Briscese:2012ys,Koshelev:2016xqb,Craps:2014wga}; while in the context of loop quantum gravity, divergences~\cite{Talaganis:2014ida,Talaganis:2015wva}; and UV finiteness~\cite{Talaganis:2017tnr} were investigated.

In GR, singularities arise because it is not possible for null rays to defocus without violating the Null Energy Condition (NEC) \cite{Hawking:1973uf}. 
Using 
the Raychaudhuri equation for null geodesic congruences~\cite{waldbook,Geroch:1968ut,Kar:2006ms}, the defocusing conditions for perturbations
around a Minkowski background were found within an IDG framework~\cite{Conroy:2016sac,Conroy:2017nkc,Conroy:2017uds}, while \cite{Conroy:2014dja} focused on a bouncing Friedmann-Robertson-Walker (FRW) model.
It was shown that IDG allows for the defocusing of null rays in that null rays can be made past-complete. 

It is this notion of geodesic completeness, or rather geodesic incompleteness, which forms the definition of a singularity for the current analysis. A photon travelling along a geodesic which is past-incomplete will simply cease to exist in a finite proper `time' (affine parameter), representative of serious weaknesses in the theory.  This is in accordance with Hawking and Ellis \cite{Hawking:1973uf}, see also \cite{waldbook, Geroch:1968ut} for discussions on the difficulties in defining singularities in GR. In this article, we will generalise the results around flat space to an (Anti) de Sitter background, where all our results 
reduce to the Minkowski values when we take the Hubble constant $H\to 0$. 

We further go on to generalise this to an (A)dS-Bianchi I metric. This metric is not isotropic as the rate of expansion differs according to direction.
Cosmological data suggests that the universe is isotropic to a very high precision, but since the observable universe makes up only 
a very small region of the universe as a whole, it is permissible that isotropy does not hold at larger scales~\cite{waldbook}.

\section{Raychaudhuri equation}
The behaviour of null geodesic congruences can be understood using the Raychaudhuri equation, which is entirely model-independent, being a purely geometric identity until it is married to a gravitational theory via the Ricci tensor. We look at $k_\mu$, a four vector 
tangential to the null geodesic congruence, where $k^\mu k_\mu=0$.
We are using a mostly positive metric signature $(-,+,+,+)$ and have defined the expansion parameter as $\theta \equiv \nabla_\mu k^\mu$.
If we assume the congruence of null rays to be orthogonal to the hypersurface, the twist tensor vanishes \cite{Vachaspati:1998dy, Kar:2006ms}. 
Additionally, since shear forces  are always positive on the right hand side of the Raychaudhuri equation, we can turn the Raychaudhuri equation into an inequality~\cite{waldbook} 
\ba \label{eq:Raych}
        \frac{d\theta}{d\lambda} + \frac{1}{2} \theta^2 \leq - R_{\mu\nu} k^\mu k^\nu,
\ea
where $\lambda$ is the affine parameter of each curve in the congruence, and $R_{\mu\nu}$ is the Ricci curvature tensor. 
In fact, the shear tensor always vanishes in FRW-type models, such as the (Anti) de Sitter universes and Minkowski space, 
as well as for a generic FRW metric~\cite{Ellis:2012}, though not necessarily in anisotropic spacetimes such as the Bianchi I models, discussed later.%can't cite section because sections don't have numbers 

Consider the inequality given by \eqref{eq:Raych}: If the right hand side is negative, we cannot have a positive and increasing expansion and therefore by the Hawking-Penrose singularity theorem~\cite{Hawking:1973uf}, a singularity results.
Therefore the condition for defocusing and thus avoiding a singularity is
\ba \label{eq:raychdefocusingconditiononriccitensor}
R_{\mu\nu} k^\mu k^\nu <0.
\ea
For a more complete treatment of this condition, see \cite{Hawking:1973uf,Raychaudhuri:1953yv,Albareti:2012va,Vachaspati:1998dy,Ellis:2012,Kar:2006ms,Vachaspati:1998dy}.  
In GR, the Einstein equation implies that the defocusing condition cannot be fulfilled, because $R_{\mu\nu} k^\mu k^\nu = \kappa T_{\mu\nu} k^\mu k^\nu$ is always positive due to the NEC 
$T_{\mu\nu} k^\mu k^\nu>0$ where $T_{\mu\nu}$ is the energy momentum tensor and $\mu$, $\nu$ run from 0 to 3.

It was shown in~\cite{Conroy:2016sac}, using modified gravity in the form of IDG that this fate can be avoided around Minkowski backgrounds. In this paper we will aim to generalise that result to (Anti) de Sitter backgrounds and a specified Bianch I model.

\section{Linearised field equations around a de Sitter background}
Conroy et al.~\cite{Conroy:2016sac} examined defocusing from perturbations around a flat background for the action~\cite{Biswas:2011ar,Biswas:2011ar} 
\ba \label{eq:action}
        S = \int d^4 x \sqrt{-g} \left( M^2_p R+ \lambda R F(\Box) R \right),
\ea
where $R$ is the Ricci scalar, $M_P$ is the Planck mass, $\lambda$ is a constant and $F(\Box)=\sum_{n=0}^\infty f_n \Box^n/M^{2n}$ 
is a function of the d'Alembertian operator $\Box$, where $\{f_0, f_1,...\}$ are the coefficients of the series.  
$M$ is the mass scale of the theory, which determines the length scales at which the extra infinite derivative terms come into play.

Laboratory tests of the departure from the $1/r$ behaviour of the Newtonian potential show that as there is no departure at $5 \times 10^{-6}$m~\cite{Kapner:2006si}, 
which gives us the constraint $M\geq 10^{-2}$eV~\cite{Edholm:2016hbt} for the simplest version of out theory. 
Analysis of the raw data from experiment 
suggests that an oscillating function given by more complicated versions of
%a more complicated version of 
IDG fits the data 
better than the GR prediction in the weak field regime~\cite{Perivolaropoulos:2016ucs,Conroy:2017nkc}.

We begin with an (A)dS metric $\bar{g}_{\mu\nu}$, background Ricci scalar $\bar{R}$
and the background Ricci tensor $\bar{R}^\alpha_\beta =\delta^\alpha_\beta \bar{R}/4$. 
We perturb around this background using $g_{\mu\nu}\to \bar{g}_{\mu\nu}+h_{\mu\nu}$, producing the linearised equations of motion  
\ba \label{eq:dsfieldeqns}
        T^\mu_\nu &=& \left( M^2_p + 2 \bar{R} \lambda f_0 \right) \left( r^\mu_\nu - \frac{1}{2} \delta^\mu_\nu r \right) \nonumber\\
        &&- 2 \lambda \left( \nabla^\mu \partial_\nu - \delta^\mu_\nu \Box \right) F(\Box) r +  \lambda \frac{\bar{R}}{2} \delta^\mu_\nu F(\Box) r. 
\ea
The Ricci tensor is perturbed as $R^\mu_\nu = \bar{R}^\mu_\nu+ r^\mu_\nu$ and the Ricci scalar is perturbed as 
$R=\bar{R}+ r$ where the linearised Ricci tensor and Ricci scalar are \cite{Biswas:2016etb,Conroy:2017uds}
\ba
        r^\mu_\nu &=& \frac{1}{2} \left( \D_\sigma \D^\mu h^\sigma_\nu 
        + \D_\nu \D_\sigma h^{\sigma\mu} - \D_\nu \D^\mu h - \Box h_\nu^\mu \right) - \frac{\bar{R}}{4} h_\nu^\mu, \nonumber\\
        r &=& \D_\mu \D_\nu h^{\mu\nu} - \Box h -\frac{\bar{R}}{4} h.
\ea   
We can write \eqref{eq:dsfieldeqns} in the neat form
\ba \label{eqn:Minkowskifieldeqns}
        M^{-2}_P T^\mu_\nu = a r^\mu_\nu - \frac{1}{2} \delta^\mu_\nu c (\Box)
r - \frac{1}{2} \nabla^\mu \partial_\nu f(\Box)r,
\ea 
similarly to the Minkowski case~\cite{Conroy:2016sac}, but where we have now defined
\ba \label{eq:defnofacf}
        && a = 1 + 2 M^{-2}_P \bar{R} \lambda f_0, \nonumber\\
        && c(\Box) = 1 + 2\lambda M^{-2}_P \bar{R} f_0 - 4 \lambda M^{-2}_P \left( \Box + \frac{\bar{R}}{3} \right) F(\Box), \nonumber\\              
        && \nabla^\mu \partial_\nu f(\Box) = 4 M^{-2}_P\lambda \left(\nabla^\mu \partial_\nu + \delta^\mu_\nu \frac{\bar{R}}{12} \right) F(\Box).
\ea
By taking the trace of the last line of (\ref{eq:defnofacf}), we can see that 
\ba    \label{eq:boxfbox}
        \Box f(\Box) = a - c(\Box),
\ea        
as for the Minkowski case~\cite{Conroy:2016sac}. 

The defocusing condition $r^\mu_\nu k_\mu k^\nu <0$ can be reached using 
the same method as~\cite{Conroy:2016sac}: 
that is, we contract (\ref{eqn:Minkowskifieldeqns}) with the null tangent vectors $k_\mu k^\nu$ where $k^\mu k_\mu=0$ 
and therefore from \eqref{eq:raychdefocusingconditiononriccitensor}, we obtain the requisite condition for defocusing 
\ba 
         r^\mu_\nu k_\mu k^\nu = \frac{1}{a} \left[ M^{-2}_P T^\mu_\nu k_\mu k^\nu +
         \frac{1}{2} k_\mu k^\nu \nabla^\mu \partial_\nu f(\Box)r\right]<0.~~~~~~
\ea
The NEC dictates that $T_{\mu\nu} k^\mu k^\nu>0$, while $a$ must be strictly
positive to avoid the introduction of ghosts~\cite{Biswas:2016etb,Biswas:2016egy} or negative entropy
states~\cite{Conroy2015wfa,Conroy:2017uds}. We may then write the minimum defocusing condition 
\ba
         k_\mu k^\nu \nabla^\mu \partial_\nu f(\Box)r<0.
\ea
If we expand the covariant derivatives
and note that if the perturbed Ricci scalar $r$ is dependent only on time, then the d'Alembertian operator is given by $\Box = - \partial^2_t - 3H \partial_t$, so that
the defocusing condition becomes\footnote{It is straightforward to verify that this reduces to the condition for perturbations around Minkowski~\cite{Conroy:2016sac}, 
by taking the limit $H\to 0$. This is equivalent to the form found in~\cite{Conroy:2017uds}.}
\ba \label{eq:defocusingcondition}
         \Box f(\Box)r(t) >- 4 H \partial_t f(\Box) r(t).
\ea
We have thus shown that it is possible to achieve defocusing around an (A)dS background. 
Therefore by using the power of IDG we can avoid the Hawking-Penrose singularity. 

\section{(A)dS-Bianchi I metric}
\label{sec:BianchiI}
\begin{figure}[h]
\includegraphics[width=9cm]{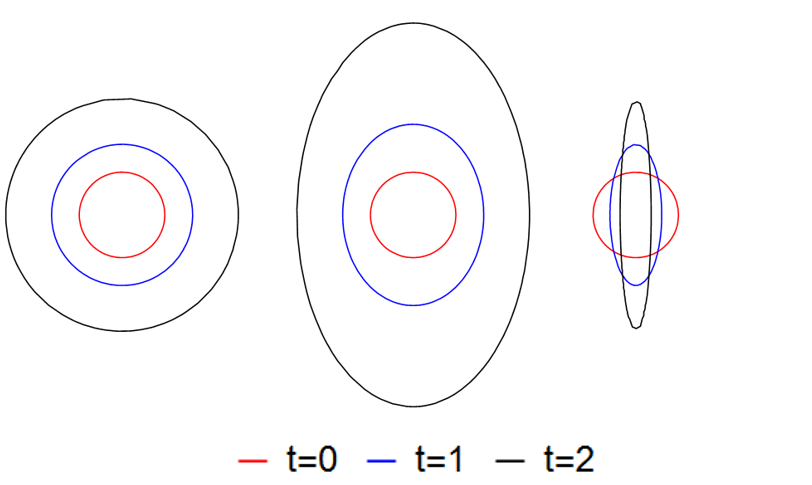} 
\caption{We plot the time evolution of three universes. The first is isotropic; the second is anisotropic and expanding in both the $x$ and $y$ directions; 
while the third is expanding in the $y$ direction but shrinking in the $x$ direction.}
\label{Fig1}
\end{figure}

Next we investigate perturbations around a more general background metric, 
where the scale factor is different in each direction. By introducing anisotropy into the system, we allow for a more general Bianchi I metric. This has the line element
\ba \label{eq:anisotropiclineelement}
        ds^2 = -dt^2 + e^{2At} dx^2 + e^{2Bt} dy^2 + e^{2Ct} dz^2, 
\ea        
which we call an (A)dS-Bianchi I metric. Examples of anisotropic universes are given in Fig.~\ref{Fig1}.
As discussed earlier, the observable universe appears to be highly isotropic~\cite{waldbook} but this may not be the case at larger scales.
A vacuum solution with a cosmological constant is only possible when $A=B=C$, i.e. when there is no anisotropy. In this paper, we assume that the positive constants $A,B$ and $C$ are roughly of the same magnitude.

The metric \eqref{eq:anisotropiclineelement} gives a constant background Ricci tensor $\bar{R}^\mu_\nu$ and positive Ricci scalar $\bar{R}=2 \left(A^2+B^2+C^2+AB+AC+B C\right)$, where bars denote background quantities. 
We perturb around the background metric, again according to $g_{\mu\nu}\to \bar{g}_{\mu\nu} +h_{\mu\nu}$. 
The equations of motion for this perturbation are  
\ba \label{eq:dsbianchieoms}
        T^\alpha_\beta &=&\left(M^2_p+2 \lambda f_0 \bar{R}\right) \left(r^\alpha_\beta
         - \frac{1}{2} \delta^\alpha_\beta
        r \right)   + 2 \lambda\bar{S}^\alpha_\beta
F(\Box)r\non
                &&+ \frac{\lambda}{2} \delta^\alpha_\beta \bar{R}F(\Box) r-2 \lambda \left(\nabla^\alpha
\nabla_\beta - \delta^\alpha_\beta \Box\right) F(\Box)r.~~~~~~~~~
\ea        
This is identical to the (A)dS field equations \eqref{eq:dsfieldeqns} with the exception of the extra term $\bar{S}^\alpha_\beta$ which vanishes for (A)dS. By redefining
\ba
        && c(\Box) = 1 + 2\lambda M^{-2}_P \bar{R} f_0 - 4 \lambda M^{-2}_P \left( \Box + \frac{\bar{R}}{12} \right) F(\Box), \nonumber\\              
        && \nabla^\mu \partial_\nu f(\Box) = 4 M^{-2}_P\lambda \left(\nabla^\mu \partial_\nu + \delta^\mu_\nu \frac{\bar{R}}{12}- \frac{1}{4}\bar{R}^\mu_\nu \right) F(\Box),~~~~~~~~
\ea   
\eqref{eq:dsbianchieoms} can be written in the neat form \eqref{eqn:Minkowskifieldeqns}. However, to see the effect of anisotropy more clearly, we continue in the form \eqref{eq:dsbianchieoms}.
Contracting \eqref{eq:dsbianchieoms} with $k^\beta k_\alpha$, where $k^\alpha k_\alpha=0$, produces the defocusing condition  
\ba
         k^\beta k_\alpha r^\alpha_\beta =&& \frac{\lambda}{M^2_P + 2 \lambda f_0 \bar{R}}\bigg(k^\beta k_\alpha T^\alpha_\beta- 2 \lambda k^\beta k_\alpha\bar{R}^\alpha_\beta
F(\Box)r\non&&+2 \lambda k^\beta k_\alpha \nabla^\alpha
\nabla_\beta F(\Box)r\bigg)<0,
\ea  
where $T^\alpha_\beta$ is the full stress energy tensor, which reduces to the (A)dS condition when we take $A=B=C=H$, and thus $\bar{R}^\alpha_\beta k_\alpha k^\beta =\frac{\bar{R}}{4}\delta^\alpha_\beta k_\alpha k^\beta=0$.
Using the NEC, the minimum defocusing condition becomes 
\ba
       \frac{\lambda}{M^2_P + 2 \lambda f_0 \bar{R}} \bigg(k^\beta k^\alpha \nabla_\alpha
\nabla_\beta  - k_\alpha  k^\beta \bar{R}^\alpha_\beta
 \bigg)F(\Box)r<0.~~~~
\ea         
We can choose
$k_0=e^{-At}$, $k_x=1$, $k_y = k_z=0$, which means that the rotation $\omega_{\alpha\beta} \equiv \nabla_\alpha k_\beta - \nabla_\beta k_\alpha$ is zero, which is required for
\eqref{eq:Raych} to be valid~\cite{Relativiststoolkit}, and the geodesic equations to be fulfilled. 
When we assume the perturbation is $t$-dependent, we find the following defocusing condition
\ba \label{eq:finaldefocusingconditionfordsbianchi}
        &&\frac{\lambda}{M^2_P + 2 \lambda f_0 \bar{R}} \bigg[ \partial_t^2
        - A \partial_t +(B^2 + C^2) \non
        &&-A (B+C)
        \bigg]F(\Box)r(t)<0,
\ea   
which reduces to the standard (A)dS case in Eq.(11) with $A=B=C=H$, as expected. Compared to the (A)dS case with $H=A$, the anisotropic (A)dS-Bianchi I metric produces an extra term in the defocusing condition:

\begin{equation}\label{eq:extra}
\frac{\lambda}{M^2_P + 2 \lambda f_0 \bar{R}}F(\Box)r(t)\left[(B^2+C^2) - A(B+C)\right].
\end{equation}
In this example we have taken the spatial part of the null ray to be in the $x$ direction. 
In order to avoid singularities it is necessary to fulfil the defocusing condition for all $k_\mu$.\footnote{If we were to replace $k_x$ with $k_y$ (or $k_z$), we would obtain an equation of the form \eqref{eq:extra} with $A\rightleftarrows B$ (or $A\rightleftarrows C$).}  
Looking first at the case where the prefactor $\frac{\lambda}{M^2_P + 2 \lambda f_0 \bar{R}}F(\Box)r(t)$ in \eqref{eq:extra} is positive (as is the case around an (A)dS background,~\cite{Conroy2015wfa}), 
$A>B,C$ implies that \eqref{eq:extra} is negative and so the anisotropy makes defocusing easier, because it
allows null geodesic congruences to more readily diverge. 
This means that the tangent vector must be in the direction of fastest expansion for anisotropy to aid defocusing. This situation is reversed for a negative prefactor. 

\section{Example of a perturbation}
Finally we choose a specific perturbation to see whether our defocusing conditions can be fulfilled. We can choose 
$h_{\mu\nu}$, where $g_{\mu\nu}\to\bar{g}_{\mu\nu}+h_{\mu\nu}$, to be $\delta^x_\mu \delta_\nu^x Pe^{2Xt}$, i.e. an exponential perturbation in the $x$-direction. Here $P$ and $X$ are dimensionless constants. 
This gives the perturbed Ricci scalar 
\ba \label{eq:perturbedr}
        r(t)=Ke^{2(X-A)t},
\ea         
where $K$ is the constant $K=2PX(2A+B+C  +2X)-2AP(B +C)$.
Therefore 
\ba \label{eq:boxperturbedr}
        \Box^n r(t) = \left[2(A-X) (2X-A+B+C)\right]^n r(t),
\ea
and since $e^{2(X-2A)t}$ is strictly positive, the defocusing condition \eqref{eq:finaldefocusingconditionfordsbianchi} becomes        
\ba  \label{eq:defocusingconditionforspecificperturbation}
        && \bigg[
4X^2-10AX+6A^2   +(B^2 + C^2)-A (B+C) \bigg]K\non
        &&\cdot\frac{ \lambda F( (A-X)(X+B+C))}{M^2_P + 2 \lambda f_0 \bar{R}}<0.
\ea         
 
Taking $y\ll A$, the term in the square brackets will be positive unless $A$ is significantly bigger than $B$ and $C$, 
which does not agree with current observations. If we take the simplest choice of the propagator $\Gamma$ which has no 
ghosts~\footnote{This choice gives the coefficients of $F(\Box)$ as $f_n= \frac{1}{6}\frac{(-1)^n}{(n+1)!}$ so $f_0=1/6$. Note that as the background Ricci scalar is positive,
the denominator of the second line of \eqref{eq:defocusingconditionforspecificperturbation} is then also strictly positive.}~\cite{Biswas:2010zk}, 
$\Gamma(\Box)=\exp(-\Box /M^2)$, and discard $X^2$ terms, the condition becomes
\ba \label{eq:finaldefocusingconditionfordsbianchiexppert}
        P> 0,
\ea        
i.e. the perturbation must be positive. It is clearly possible to fulfil the defocusing condition and avoid the singularities which plague the theory of General Relativity.
%%%%%%%%%%%%%%%%%%%%%%%%%%%%%%%%%%%%%%%%%%%%%%%%%%%%%%%%%%%%%%%%%%%%%%%%%%%%%%%%%%%%%%%%%%%%%%%%%%%%%%%%%%%%%%%%%%%%%%%%%%%%%%%%%%%%%%%%%%%%%%%%%%%%%%
\section{Conclusion}
In this short note, we have shown that it is possible for null rays to defocus in an (Anti) de Sitter background, using ghost-free Infinite Derivative Gravity.     
This allowed us to avoid the initial singularity that naturally occurs in General Relativity according to the Hawking-Penrose Singularity Theorem. 

Furthermore, we extended the analysis to an anisotropic and homogenous Bianchi I model resembling an anistropic ``(Anti) de Sitter space''. Finally, we looked at a particular perturbation within this model and showed that defocusing
was possible given certain constraints on the parameters of the metric, the perturbation and the form of $F(\Box)$.

%%%%%%%%%%%%%%%%%%%%%%%%%%%%%%%%%%%%%%%%%%%%%%%%%%%%%%%%%%%%%%%%%%%%%%%%%%%%%%%%%%%%%%%%%%%%%%%%%%%%%%%%%%%%%%%%%%%%%%%%%%%%%%%%%%%%%%%%%%%%%%%%%%%%%%
\section{Acknowledgements}
We would like to thank David Burton and Charlotte Owen for their advice and suggestions in preparing this paper.

%%%%%%%%%%%%%%%%%%%%%%%%%%%%%%%%%%%%%%%%%%%%%%%%%%%%%%%%%%%%%%%%%%%%%%%%%%%%%%%%%%%%%%%%%%%%%%%%%%%%%%%%%%%%%%%%%%%%%%%%%%%%%%%%%%%%%%%%%%%%%%%%%%%%%%
%%%%%%%%%%%%%%%%%%%%%%%%%%%%%%%%%%%%%%%%%%%%%%%%%%%%%%%%%%%%%%%%%%%%%%%%%%%%%%%%%%%%%%%%%%%%%%%%%%%%%%%%%%%%%%%%%%%%%%%%%%%%%%%%%%%%%%%%%%%%%%%%%%%%%%%    
%+Bibliography

%-Bibliography


\begin{thebibliography}{99}

\bibitem{Will}
C. M. Will, Living Rev. Relativity, 17, (2014), 4
``The Confrontation between General Relativity and Experiment,''
doi:10.12942/lrr-2014-4
\bibitem{Hawking:1973uf}
  S.~W.~Hawking and G.~F.~R.~Ellis,
  ``The Large Scale Structure of Space-Time,''
  doi:10.1017/CBO9780511524646
  %%CITATION = doi:10.1017/CBO9780511524646;%%


%\cite{Raychaudhuri:1953yv}
\bibitem{Raychaudhuri:1953yv}
A.~Raychaudhuri,
``Relativistic cosmology. 1.,''
Phys.\ Rev.\  {\bf 98} (1955) 1123.
doi:10.1103/PhysRev.98.1123
%%CITATION = doi:10.1103/PhysRev.98.1123;%%
%214 citations counted in INSPIRE as of 17 Jul 2017
  
\bibitem{Stelle:1976gc}
  K.~S.~Stelle,
  ``Renormalization of Higher Derivative Quantum Gravity,''
  Phys.\ Rev.\ D {\bf 16} (1977) 953.
  doi:10.1103/PhysRevD.16.953
  %%CITATION = doi:10.1103/PhysRevD.16.953;%%
  %1252 citations counted in INSPIRE as of 07 Aug 2016
      
\bibitem{Barnaby:2007ve}
  N.~Barnaby and N.~Kamran,
  ``Dynamics with infinitely many derivatives: The Initial value problem,''
  JHEP {\bf 0802} (2008) 008
  doi:10.1088/1126-6708/2008/02/008
  [arXiv:0709.3968 [hep-th]].
  %%CITATION = doi:10.1088/1126-6708/2008/02/008;%% 
       
\bibitem{VanNieuwenhuizen:1973fi}
  P.~Van Nieuwenhuizen,
  ``On ghost-free tensor lagrangians and linearized gravitation,''
  Nucl.\ Phys.\ B {\bf 60} (1973) 478.
  doi:10.1016/0550-3213(73)90194-6
         
 \bibitem{Woodard:2015zca}
 R.~P.~Woodard,
 ``Ostrogradsky's theorem on Hamiltonian instability,''
 Scholarpedia {\bf 10} (2015) no.8,  32243
 doi:10.4249/scholarpedia.32243
 [arXiv:1506.02210 [hep-th]].
 %%CITATION = doi:10.4249/scholarpedia.32243;%%
 
 \bibitem{Ostrogradsky:1850fid}
 M.~Ostrogradsky,
 ``Mémoires sur les équations différentielles, relatives au problème des isopérimètres,''
 Mem.\ Acad.\ St.\ Petersbourg {\bf 6} (1850) no.4,  385.
      
\bibitem{Tseytlin:1995uq}
  A.~A.~Tseytlin,
  ``On singularities of spherically symmetric backgrounds in string theory,''
  Phys.\ Lett.\ B {\bf 363} (1995) 223
  doi:10.1016/0370-2693(95)01228-7
  [hep-th/9509050].
        
\bibitem{Biswas:2005qr}
  T.~Biswas, A.~Mazumdar and W.~Siegel,
  ``Bouncing universes in string-inspired gravity,''
  JCAP {\bf 0603} (2006) 009
  doi:10.1088/1475-7516/2006/03/009
  [hep-th/0508194].
  %%CITATION = doi:10.1088/1475-7516/2006/03/009;%%
          
\bibitem{Biswas:2011ar}
  T.~Biswas, E.~Gerwick, T.~Koivisto and A.~Mazumdar,
  ``Towards singularity and ghost free theories of gravity,''
  Phys.\ Rev.\ Lett.\  {\bf 108} (2012) 031101
  doi:10.1103/PhysRevLett.108.031101
  [arXiv:1110.5249 [gr-qc]].
  %%CITATION = doi:10.1103/PhysRevLett.108.031101;%%

\bibitem{Biswas:2016etb}
  T.~Biswas, A.~S.~Koshelev and A.~Mazumdar,
  ``Gravitational theories with stable (anti-)de Sitter backgrounds,''
  Fundam.\ Theor.\ Phys.\  {\bf 183} (2016) 97
  doi:10.1007/978-3-319-31299-6
  [arXiv:1602.08475 [hep-th]].
  %%CITATION = doi:10.1007/978-3-319-31299-6_5;%% 
              
\bibitem{Tomboulis}
   E.~Tomboulis,
  ``Renormalizability and Asymptotic Freedom in Quantum Gravity,''
 {\it Phys. Lett. B} {\bf 97}, 77 (1980).
  %%CITATION = PHLTA,B97,77;%%      
   
\bibitem{Siegel:2003vt} 
  W.~Siegel,
  ``Stringy gravity at short distances,''
  hep-th/0309093.
  %%CITATION = HEP-TH/0309093;%% 
  
\bibitem{Biswas:2013kla} 
  T.~Biswas, T.~Koivisto and A.~Mazumdar,
  ``Nonlocal theories of gravity: the flat space propagator,''
  arXiv:1302.0532 [gr-qc].
  %%CITATION = ARXIV:1302.0532;%%
           
          
\bibitem{Buoninfante:2016iuf}
  L.~Buoninfante,
  ``Ghost and singularity free theories of gravity,''
  arXiv:1610.08744 [gr-qc].
  %%CITATION = ARXIV:1610.08744;%%        
  
\bibitem{Talaganis:2014ida}
  S.~Talaganis, T.~Biswas and A.~Mazumdar,
  ``Towards understanding the ultraviolet behavior of quantum loops in infinite-derivative
theories of gravity,''
  Class.\ Quant.\ Grav.\  {\bf 32} (2015) no.21,  215017
  doi:10.1088/0264-9381/32/21/215017
  [arXiv:1412.3467 [hep-th]].
  %%CITATION = doi:10.1088/0264-9381/32/21/215017;%% 
         
\bibitem{Modesto:2011kw}
L.~Modesto,
``Super-renormalizable Quantum Gravity,''
  Phys.\ Rev.\ D {\bf 86} (2012) 044005
  doi:10.1103/PhysRevD.86.044005
  [arXiv:1107.2403 [hep-th]].

\bibitem{Modesto:2012ys}
  L.~Modesto,
  ``Super-renormalizable Multidimensional Quantum Gravity,''
  Astron. Rev. 8.2 (2013) 4-33
  [arXiv:1202.3151 [hep-th]].
             
\bibitem{Modesto:2014lga}
  L.~Modesto and L.~Rachwal,
  ``Super-renormalizable and finite gravitational theories,''
  Nucl.\ Phys.\ B {\bf 889} (2014) 228
  doi:10.1016/j.nuclphysb.2014.10.015
  [arXiv:1407.8036 [hep-th]].
            
\bibitem{Modesto:2016max}
  L.~Modesto and L.~Rachwal,
  ``Finite Conformal Quantum Gravity and Nonsingular Spacetimes,''
  arXiv:1605.04173 [hep-th].
                 
\bibitem{Edholm:2016hbt}
  J.~Edholm, A.~S.~Koshelev and A.~Mazumdar,
  ``Universality of testing ghost-free gravity,''
  arXiv:1604.01989 [gr-qc].
  Phys.\ Rev.\ D {\bf 94} (2016) no.10,  104033
  doi:10.1103/PhysRevD.94.104033
  %%CITATION = ARXIV:1604.01989;%%    

\bibitem{Conroy:2014eja}
  A.~Conroy, T.~Koivisto, A.~Mazumdar and A.~Teimouri,
  ``Generalized quadratic curvature, non-local infrared modifications of gravity and Newtonian potentials,''
  Class.\ Quant.\ Grav.\  {\bf 32} (2015) no.1,  015024
  doi:10.1088/0264-9381/32/1/015024
  [arXiv:1406.4998 [hep-th]].
  %%CITATION = doi:10.1088/0264-9381/32/1/015024;%%
  
\bibitem{Conroy:2017nkc}
  A.~Conroy and J.~Edholm,
  ``Newtonian Potential and Geodesic Completeness in Infinite Derivative Gravity,''
  arXiv:1705.02382 [gr-qc].
  %%CITATION = ARXIV:1705.02382;%%
  
\bibitem{Frolov}  
 V.~P.~Frolov and A.~Zelnikov,
  ``Head-on collision of ultra-relativistic particles in ghost-free theories of gravity,''
  arXiv:1509.03336 [hep-th].
  %%CITATION = ARXIV:1509.03336;%%
  
\bibitem{Frolov2015mfb}
V.~P.~Frolov and A.~Zelnikov,
  ``Mass-gap for black hole formation in higher derivative and ghost free gravity,''
  {\it Phys. Rev. Lett.}  {\bf 115}, no. 5, 051102 (2015)
  [arXiv:1505.00492 [hep-th]].
  %%CITATION = doi:10.1103/PhysRevLett.115.051102;%%
  %7 citations counted in INSPIRE as of 16 Dec 2015
  
\bibitem{Frolov2015scs}
V.~P.~Frolov, A.~Zelnikov and T.~de Paula Netto,
  ``Spherical collapse of small masses in the ghost-free gravity,''
  {\it JHEP} {\bf 1506}, 107 (2015)
  [arXiv:1504.00412 [hep-th]].
  %%CITATION = doi:10.1007/JHEP06(2015)107;%% 

\bibitem{Biswas:2013cha}
  T.~Biswas, A.~Conroy, A.~S.~Koshelev and A.~Mazumdar,
  ``Generalized ghost-free quadratic curvature gravity,''
  Class.\ Quant.\ Grav.\  {\bf 31} (2014) 015022
   Erratum: [Class.\ Quant.\ Grav.\  {\bf 31} (2014) 159501]
  doi:10.1088/0264-9381/31/1/015022, 10.1088/0264-9381/31/15/159501
  [arXiv:1308.2319 [hep-th]].
   
\bibitem{Calcagni:2013vra}
G.~Calcagni, L.~Modesto and P.~Nicolini,
``Super-accelerating bouncing cosmology in asymptotically-free non-local gravity,''
  Eur.\ Phys.\ J.\ C {\bf 74} (2014) no.8,  2999
  doi:10.1140/epjc/s10052-014-2999-8
  [arXiv:1306.5332 [gr-qc]].
     
\bibitem{Biswas:2012bp}
  T.~Biswas, A.~S.~Koshelev, A.~Mazumdar and S.~Y.~Vernov,
  ``Stable bounce and inflation in non-local higher derivative cosmology,''
  JCAP {\bf 1208} (2012) 024
  doi:10.1088/1475-7516/2012/08/024
  [arXiv:1206.6374 [astro-ph.CO]].
  %%CITATION = doi:10.1088/1475-7516/2012/08/024;%%

\bibitem{Biswas:2010zk}
  T.~Biswas, T.~Koivisto and A.~Mazumdar,
  ``Towards a resolution of the cosmological singularity in non-local higher derivative theories of gravity,''
  JCAP {\bf 1011} (2010) 008
  doi:10.1088/1475-7516/2010/11/008
  [arXiv:1005.0590 [hep-th]]. 
  
  \bibitem{Koshelev:2012qn}
  A.~S.~Koshelev and S.~Y.~Vernov,
  ``On bouncing solutions in non-local gravity,''
  Phys.\ Part.\ Nucl.\  {\bf 43} (2012) 666
  doi:10.1134/S106377961205019X
  [arXiv:1202.1289 [hep-th]].
  %%CITATION = doi:10.1134/S106377961205019X;%%
  %51 citations counted in INSPIRE as of 02 Jun 2017
  
\bibitem{Koshelev:2013lfm}
  A.~S.~Koshelev,
  ``Stable analytic bounce in non-local Einstein-Gauss-Bonnet cosmology,''
  Class.\ Quant.\ Grav.\  {\bf 30} (2013) 155001
  doi:10.1088/0264-9381/30/15/155001
  [arXiv:1302.2140 [astro-ph.CO]].
  %%CITATION = doi:10.1088/0264-9381/30/15/155001;%%
  %34 citations counted in INSPIRE as of 04 Jun 2017
  
  \bibitem{Biswas:2016egy}
  T.~Biswas, A.~S.~Koshelev and A.~Mazumdar,
  ``Consistent Higher Derivative Gravitational theories with stable de Sitter and Anti-de Sitter Backgrounds,''
  arXiv:1606.01250 [gr-qc].
  %%CITATION = ARXIV:1606.01250;%%
   
% \bibitem{Conroy:2015wfa} 
%  A.~Conroy, A.~Mazumdar and A.~Teimouri,
%  ``Wald Entropy for Ghost-Free, Infinite Derivative Theories of Gravity,''
%  Phys.\ Rev.\ Lett.\  {\bf 114}, no. 20, 201101 (2015)
%  doi:10.1103/PhysRevLett.114.201101
%  [arXiv:1503.05568 [hep-th]].
%  %%CITATION = doi:10.1103/PhysRevLett.114.201101;%%
%  %10 citations counted in INSPIRE as of 21 May 2016
 %\cite{Conroy:2015nva}

 
\bibitem{Conroy2015wfa}  
A.~Conroy, A.~Mazumdar, S.~Talaganis and A.~Teimouri,
  ``Nonlocal gravity in D dimensions: Propagators, entropy, and a bouncing cosmology,''
  Phys.\ Rev.\ D {\bf 92}, no. 12, 124051 (2015)
  doi:10.1103/PhysRevD.92.124051
  [arXiv:1509.01247 [hep-th]].
  %%CITATION = doi:10.1103/PhysRevD.92.124051;%%



\bibitem{Teimouri:2016ulk}
  A.~Teimouri, S.~Talaganis, J.~Edholm and A.~Mazumdar,
  ``Generalised Boundary for Higher Derivative Theories of Gravity,''
  arXiv:1606.01911 [gr-qc].
  JHEP {\bf 1608} (2016) 144
  doi:10.1007/JHEP08(2016)144
  %%CITATION = ARXIV:1606.01911;%%
  
  \bibitem{Mazumdar:2017kxr}
  S.~Talaganis and A.~Teimouri,
  ``Hamiltonian Analysis for Infinite Derivative Field Theories and Gravity,''
  arXiv:1701.01009 [hep-th].
  %%CITATION = ARXIV:1701.01009;%%
  %5 citations counted in INSPIRE as of 30 May 2017
   
\bibitem{Modesto:2010uh}
  L.~Modesto, J.~W.~Moffat and P.~Nicolini,
  ``Black holes in an ultraviolet complete quantum gravity,''
  Phys.\ Lett.\ B {\bf 695} (2011) 397
  doi:10.1016/j.physletb.2010.11.046
  [arXiv:1010.0680 [gr-qc]].
     
\bibitem{Frolov:2016xhq}
  V.~P.~Frolov and A.~Zelnikov,
  ``Radiation from an emitter in the ghost free scalar theory,''
  Phys.\ Rev.\ D {\bf 93} (2016) no.10,  105048
  doi:10.1103/PhysRevD.93.105048
  [arXiv:1603.00826 [hep-th]].
       
\bibitem{Calcagni:2014vxa}
  G.~Calcagni and L.~Modesto,
  ``Nonlocal quantum gravity and M-theory,''
  Phys.\ Rev.\ D {\bf 91} (2015) no.12,  124059
  doi:10.1103/PhysRevD.91.124059
  [arXiv:1404.2137 [hep-th]].
  
 
     
  \bibitem{Feng:2017vqd}
  L.~Feng,
  ``Light Bending in the Infinite Derivative Theories of Gravity,''
  arXiv:1703.06535 [gr-qc].
  %%CITATION = ARXIV:1703.06535;%%  

\bibitem{Calcagni:2010ab}
  G.~Calcagni and G.~Nardelli,
  %
  ``Non-local gravity and the diffusion equation,''
  Phys.\ Rev.\ D {\bf 82} (2010) 123518
  doi:10.1103/PhysRevD.82.123518
  [arXiv:1004.5144 [hep-th]].
  
\bibitem{Koshelev:2017bxd}
  A.~S.~Koshelev and A.~Mazumdar,
  ``Absence of event horizon in massive compact objects in infinite derivative gravity,''
 arXiv:1707.00273 [gr-qc].
     
\bibitem{Calcagni:2017sov}
  G.~Calcagni and L.~Modesto,
  ``Stability of Schwarzschild singularity in non-local gravity,''
  Phys.\ Lett.\ B {\bf 773} (2017) 596
  doi:10.1016/j.physletb.2017.09.018
  
\bibitem{ArkaniHamed:2002fu}
N.~Arkani-Hamed, S.~Dimopoulos, G.~Dvali and G.~Gabadadze,
``Nonlocal modification of gravity and the cosmological constant problem,''
hep-th/0209227.

\bibitem{Edholm:2016seu}
  J.~Edholm,
  ``UV completion of the Starobinsky model, tensor-to-scalar ratio, and constraints on nonlocality,''
  Phys.\ Rev.\ D {\bf 95} (2017) no.4,  044004
  doi:10.1103/PhysRevD.95.044004
  [arXiv:1611.05062 [gr-qc]].
  %%CITATION = doi:10.1103/PhysRevD.95.044004;%%
  

  
\bibitem{Briscese:2012ys}
  F.~Briscese, A.~Marcianò, L.~Modesto and E.~N.~Saridakis,
  ``Inflation in (Super-)renormalizable Gravity,''
  Phys.\ Rev.\ D {\bf 87} (2013) no.8,  083507
  doi:10.1103/PhysRevD.87.083507
  [arXiv:1212.3611 [hep-th]].
  %%CITATION = doi:10.1103/PhysRevD.87.083507;%%
  %47 citations counted in INSPIRE as of 15 May 2017


\bibitem{Koshelev:2016xqb}
  A.~S.~Koshelev, L.~Modesto, L.~Rachwal and A.~A.~Starobinsky,
  ``Occurrence of exact $R^2$ inflation in non-local UV-complete gravity,''
  JHEP {\bf 1611} (2016) 067
  doi:10.1007/JHEP11(2016)067
  [arXiv:1604.03127 [hep-th]].
  %%CITATION = doi:10.1007/JHEP11(2016)067;%%
  
\bibitem{Craps:2014wga}
  B.~Craps, T.~De Jonckheere and A.~S.~Koshelev,
  ``Cosmological perturbations in non-local higher-derivative gravity,''
  JCAP {\bf 1411} (2014) no.11,  022
  doi:10.1088/1475-7516/2014/11/022
  [arXiv:1407.4982 [hep-th]].
 
  \bibitem{Talaganis:2015wva}
  S.~Talaganis,
  ``Quantum Loops in Non-Local Gravity,''
  PoS CORFU {\bf 2014} (2015) 162
  [arXiv:1508.07410 [hep-th]].
  %%CITATION = ARXIV:1508.07410;%%
 
  \bibitem{Talaganis:2017tnr}
  S.~Talaganis,
  ``Towards UV Finiteness of Infinite Derivative Theories of Gravity and Field Theories,''
  arXiv:1704.08674 [hep-th].
  %%CITATION = ARXIV:1704.08674;%%
  %3 citations counted in INSPIRE as of 30 May 2017
  
\bibitem{Geroch:1968ut}
R.~P.~Geroch,
``What is a singularity in general relativity?,''
Annals Phys.\  {\bf 48} (1968) 526.
doi:10.1016/0003-4916(68)90144-9
%%CITATION = doi:10.1016/0003-4916(68)90144-9;%%
%117 citations counted in INSPIRE as of 27 Sep 2017

%\cite{Kar:2006ms}
\bibitem{Kar:2006ms}
S.~Kar and S.~SenGupta,
``The Raychaudhuri equations: A Brief review,''
Pramana {\bf 69} (2007) 49
doi:10.1007/s12043-007-0110-9
[gr-qc/0611123].
%%CITATION = doi:10.1007/s12043-007-0110-9;%%
%62 citations counted in INSPIRE as of 27 Sep 2017
  
 \bibitem{waldbook}
Robert M. Wald,
``General Relativity,''
Chicago, Usa: Univ.
Pr. (1984) 

\bibitem{Conroy:2016sac}
  A.~Conroy, A.~S.~Koshelev and A.~Mazumdar,
  ``Criteria for resolving the cosmological singularity in Infinite Derivative Gravity,''
  arXiv:1605.02080 [gr-qc].
  %%CITATION = ARXIV:1605.02080;%%
  
\bibitem{Conroy:2017uds}
  A.~Conroy,
  ``Infinite Derivative Gravity: A Ghost and Singularity-free Theory,''
  arXiv:1704.07211 [gr-qc].
  %%CITATION = ARXIV:1704.07211;%%
  
    
  \bibitem{Conroy:2014dja}
  A.~Conroy, A.~S.~Koshelev and A.~Mazumdar,
  ``Geodesic completeness and homogeneity condition for cosmic inflation,''
  Phys.\ Rev.\ D {\bf 90} (2014) no.12,  123525
  doi:10.1103/PhysRevD.90.123525
  [arXiv:1408.6205 [gr-qc]].
  %%CITATION = doi:10.1103/PhysRevD.90.123525;%%
  %14 citations counted in INSPIRE as of 26 Apr 2017 
  

\bibitem{Vachaspati:1998dy}
  T.~Vachaspati and M.~Trodden,
  ``Causality and cosmic inflation,''
  Phys.\ Rev.\ D {\bf 61} (1999) 023502
  doi:10.1103/PhysRevD.61.023502
  [gr-qc/9811037].
  
 \bibitem{Ellis:2012}
 G.~F.~R.~Ellis, R.~Maartens, M.~A.~H.~MacCallum,
 ``Relativistic Cosmology,''
 Cambridge University Press
  
\bibitem{Albareti:2012va}
  F.~D.~Albareti, J.~A.~R.~Cembranos, A.~de la Cruz-Dombriz and A.~Dobado,
  ``On the non-attractive character of gravity in f(R) theories,''
  JCAP {\bf 1307} (2013) 009
  doi:10.1088/1475-7516/2013/07/009
  [arXiv:1212.4781 [gr-qc]].
  %%CITATION = doi:10.1088/1475-7516/2013/07/009;%%
    
\bibitem{Kapner:2006si}
  D.~J.~Kapner, T.~S.~Cook, E.~G.~Adelberger, J.~H.~Gundlach, B.~R.~Heckel,
C.~D.~Hoyle and H.~E.~Swanson,
  ``Tests of the gravitational inverse-square law below the dark-energy
length scale,''
  Phys.\ Rev.\ Lett.\  {\bf 98} (2007) 021101
  %doi:10.1103/PhysRevLett.98.021101
 % [hep-ph/0611184].

\bibitem{Perivolaropoulos:2016ucs}
  L.~Perivolaropoulos,
  ``Sub-millimeter Spatial Oscillations of Newton's Constant: Theoretical
Models and Laboratory Tests,''
  arXiv:1611.07293 [gr-qc].

\bibitem{Relativiststoolkit}
 Eric Poisson,
 ``A Relativist's Toolkit: The Mathematics of Black-Hole Mechanics,''  
 Cambridge University Press, 2004
 

\end{thebibliography}
\end{document}